# Accepted Manuscript

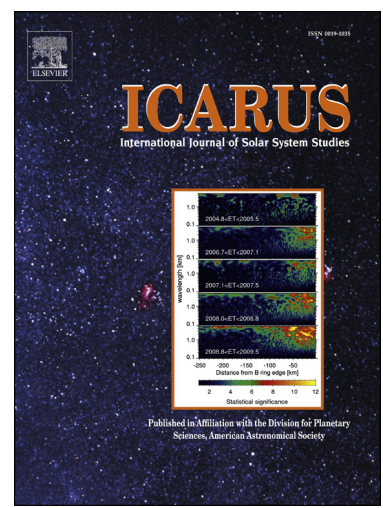

Photochemical Escape of Oxygen from Early Mars

Jinjin Zhao, Feng Tian

# Photochemical Escape of Oxygen from Early Mars

Jinjin Zhao and Feng Tian[*]

Ministry of Education Key Laboratory for Earth System Modeling, Center for Earth

System Science, Tsinghua University, Beijing, China, 100084

---

[*] Corresponding author.

E-mail address: tianfengco@tsinghua.edu.cn.

**Abstract**

Photochemical escape is an important process for oxygen escape from present Mars. In this work, a 1-D Monte-Carlo Model is developed to calculate escape rates of energetic oxygen atoms produced from $O_2^+$ dissociative recombination reactions (DR) under 1, 3, 10, and 20 times present solar XUV fluxes. We found that although the overall DR rates increase with solar XUV flux almost linearly, oxygen escape rate increases from 1× to 10× present solar XUV conditions but decreases when increasing solar XUV flux further. Analysis shows that atomic species in the upper thermosphere of early Mars increases more rapidly than $O_2^+$ when increasing XUV fluxes. While the latter is the source of energetic O atoms, the former increases the collision probability and thus decreases the escape probability of energetic O. Our results suggest that photochemical escape be a less important escape mechanism than previously thought for the loss of water and/or $CO_2$ from early Mars.

*Key words: Atmospheres, evolution; Mars, atmosphere; Photochemistry*

## 1 Introduction

Atmosphere escape is important for the evolution of Mars and could have contributed to its long term climate change (Tian et al., 2013). On present Mars, the exobase temperature is too low for thermal escape of heavy atoms, such as oxygen, to be efficient (Gröller et al., 2014; Johnson et al., 2008; Lammer et al., 2008). Nonthermal escape processes do not depend on exobase temperature and rely on sources of energetic atoms or ions. Dissociative recombination (DR) reaction of $O_2^+$ (see the following list) is one such source (Fox and Hać, 2009):

Branching Ratio

(A) $\quad O_2^+ + e^- \rightarrow O(^3P) + O(^3P) + 6.99 eV \quad\quad$ 26.5%

(B) $\quad O_2^+ + e^- \rightarrow O(^1D) + O(^3P) + 5.02 eV \quad\quad$ 47.3%

(C) $\quad O_2^+ + e^- \rightarrow O(^1D) + O(^1D) + 3.06 eV \quad\quad$ 20.4%

(D) $\quad O_2^+ + e^- \rightarrow O(^1D) + O(^1S) + 0.83 eV \quad\quad$ 5.8%

Because of the low mass of Mars (escape energy of 1.9 eV, 9.7 eV and 8.6 eV are needed for oxygen atoms to escape from Mars, Earth and Venus respectively) (Shizgal and Arkos, 1996), energetic oxygen atoms produced from branches A, B, and C can escape from the planet. This type of photochemical escape is potentially important for the evolution of volatile inventories on Mars (Lammer et al., 2003).

Because energetic O atoms are produced mainly in the collisional regions of Martian upper atmosphere, these particles need to overcome collisions with background gases in order to escape. Many workers have studied photochemical escape of oxygen from present Mars (Chassefière et al., 2013; Cipriani et al., 2007; Fox, 1993; Fox and Hać, 2009; Fox and Hać, 2014; Gröller et al., 2014; Hodges, 2000; Kim et al., 1998;

Krestyanikova and Shematovich, 2006; Lammer and Bauer, 1991; Shematovich, 2013; Valeille et al., 2010; Yagi et al., 2012; Zhang et al., 1993).

Fox and Hać (2009) considered isotropic and forward scattering in the center-of-mass frame in a 3D Monte Carlo model. A constant total cross section of $3\times10^{-15}$ cm$^2$ and differential cross sections from Kharchenko et al. (2000) are used for O-CO$_2$, O-CO, and O-O collisions. The escape rates in their forward scattering model are $1.44\times10^{26}$ s$^{-1}$ and $2.1\times10^{26}$ s$^{-1}$ for low and high solar activity situations respectively, one order of magnitude greater than those in the isotropic model (Fox and Hać, 2009). The small difference between the low and high solar activity in the forward scattering model is caused by the opposite changing trends of O$_2^+$ DR rate and escape probability of energetic oxygen in different solar activity situation (Fox and Hać, 2009).

Valeille et al. (2010) considered different solar zenith angle in a 2-D Direct Simulation Monte Carlo (DSMC) model and obtained oxygen escape rate between $3.8\sim14\times10^{25}$ s$^{-1}$. It is also shown that the oxygen escape rates change from $3\times10^{25}$ to $5\times10^{25}$ s$^{-1}$ between different seasons of Mars (Yagi et al., 2012). Shematovich (2013) considered the collisions between energetic oxygen and hydrogen and calculated oxygen escape rate of $1.5\times10^{25}$ s$^{-1}$.

More recently Fox and Hać (2014) updated their model with larger

O-CO$_2$ and O-CO collision cross sections (2.0×10$^{-14}$ and 1.8×10$^{-14}$ cm$^2$ respectively) and energy dependent on O-O collision cross section which used by Kharchenko et al. (2000). Because of CO$_2$ is the main component of modern Martian atmosphere below about 200 km altitude, new collision cross sections decrease oxygen photochemical escape rates to 6.5×10$^{24}$ s$^{-1}$ and 1.6×10$^{25}$ s$^{-1}$ for low and high solar activity situations respectively (Fox and Hać, 2014).

Another recent work including DR of O$_2^+$ and CO$_2^+$ in a 1-D Monte Carlo Model (Gröller et al., 2014) found oxygen escape rates at exobase from DR of O$_2^+$ to be 1.5×10$^{25}$ s$^{-1}$ and 2.1×10$^{25}$ s$^{-1}$ for low and high solar activity conditions respectively. In comparison, oxygen escape rates at exobase from DR of CO$_2^+$ are found to be 1.3×10$^{25}$ and 1.1×10$^{25}$ s$^{-1}$ for low and high solar activity conditions respectively (Gröller et al., 2014). Note that DR of CO$_2^+$ has been deemed unimportent in most previous works on photochemical escape of oxygen.

Despite the differences between models, the calculated photochemical escape rates of oxygen are on the order of 10$^{25}$ s$^{-1}$ for present Mars. In comparison, there are fewer works discussing photochemical escape of O atoms from early Mars. The pioneer work of Zhang et al. (1993) suggested that photochemical oxygen escape rate from ancient Martian atmosphere under 1, 3 and 6 times present solar XUV flux were 7×10$^{25}$ s$^{-1}$, 4×10$^{26}$ s$^{-1}$ and 1×10$^{27}$ s$^{-1}$ respectively (Zhang et al., 1993). Lammer et al. (2003) scaled oxygen DR escape rate of $6\times10^{24}$ s$^{-1}$ for present Mars to $3\times10^{25}$ s$^{-1}$ at 2.5 Ga (3× XUV) and $8\times10^{25}$ s$^{-1}$ at 3.5 Ga (6× XUV) respectively. It is a general consensus that oxygen photochemical escape rate from Mars should increase monotonically with increasing solar XUV.

Tian et al. (2009) proposed that the upper atmosphere of early Mars was highly expanded (exobase moves to a few Martian radii) during the early Noachian and that a $CO_2$ atmosphere of early Mars was unstable against solar XUV driven thermal escape (Tian et al., 2009). No previous work has modeled photochemical escape in such a scenario.

In this work a 1-D Monte Carlo model is used to investigate the oxygen photochemical escape from a highly expanded atmosphere. We found that photochemical escape of oxygen from early Martian atmosphere does not increase with solar XUV monotonically. Instead there is a threshold beyond which the expansion of the upper atmosphere, mainly consists of atomic species, exceed the increase of $O_2^+$ and thus escape rate of energetic O atoms decreases with increasing XUV beyond that threshold. The next section describes the model. Section 3 is the results and discussion for early Mars.

## 2 Model Description

The neutral and electron temperature profiles, as well as the density profiles of O, CO, $CO_2$, $O_2^+$, and $e^-$, used in this work (shown in Fig.1 and Fig.2) under 1×, 3×, 10× and 20× XUV conditions are based on Tian et al. (2009). The top of the upper atmosphere are set to 300 km, 500 km, 1200 km and 10,000 km respectively. At the altitudes for which Tian et al. (2009) model does not provide data, densities and temperatures are extrapolated. The rate coefficient for the $O_2^+$ DR reaction is set to $1.95 \times 10^{-7} (300/T_e)^{0.7}$ cm$^3$ s$^{-1}$ for $T_e < 1200$ K and $7.39 \times 10^{-8} (1200/T_e)^{0.56}$ cm$^3$ s$^{-1}$ for $T_e > 1200$ K (Fox and Hać, 2009).

To calculate escape probabilities of energetic oxygen atoms formed at different altitudes with different energies, $6 \times 10^7$ test particles, initial kinetic energy uniformly distribute from 0.03 eV to 10.02 eV with 0.03 eV bin size, are launched at each altitude grid. Random collisions between 3 type collision pairs, O-O, O-$CO_2$ and O-CO, are considered in the simulation. The occurrence of collisions and the target species with which energetic oxygen should collide with are calculated based on the collision probability $P_i$:

$$P_i = 1 - \exp(-\sigma_i n_i \Delta h)$$

Where $n_i$ is the number density of background species i, $\Delta h$ is the distance the particle traveled, set to be the smaller of 20% of local mean free path and 1 km (Fox and Hać, 2009), $\sigma_i$ is the collision cross section of the energetic O atoms with background species i. For O-$CO_2$ and O-CO collisions, the collision cross sections are $2.0 \times 10^{-14}$ cm$^2$ and $1.8\times10^{-14}$ cm$^2$ respectively (Fox and Hać (2014). For O-O collisions, the energy dependent diffusion collision cross sections and differential cross sections in Tully and Johnson (2001) are used. When collisions occur, the energy loss of the projectile is calculated based on the diffusion cross sections in Tully and Johnson (2001) for O-O collisions. For O-CO$_2$ and O-CO collisions, we assume 100% energy loss for simplicity.

For each test particle, its kinetic energy after each collision is compared with the escape energy of Mars. Particles reaching the top of the atmosphere and still possess kinetic energy exceeding the escape energy contribute to the calculations of escape probability of oxygen atoms from particular altitudes.

Secondary energetic oxygen atoms, background oxygen atoms gaining energy through collisions from the primary energetic oxygen atoms formed directly from the DR reactions, could escape from Mars if their kinetic energy exceeds the local escape energy. To account for them, the model stores the number of the secondary test particles capable of escaping (their initial energy be identical to the energy lost by the parent primary particles), the altitude where they are formed, and the altitude where their parent test particles were launched. Then a separate model, similar to that of the primary test particles, is carried out in which the sum of escaping primary test particles launched from one particular altitude and the escaping secondary test particles produced by primary test particles launched from the same altitude is used to calculate the escape probability of oxygen atoms from that particular altitude.

The altitude dependence of the escape energy of oxygen atoms and the reduction of the particle's kinetic energy in Mars' gravity field are considered in the model. The escape rates are calculated based on the production rates of energetic O atoms and the escape probabilities at different altitudes.

Fox and Hać (2009) discussed the contributions of DR from vibrationally excited $O_2^+$, which are from the following reactions:

$$O^+ + CO_2 \rightarrow O_2^+(v) + CO + 1.19 \text{eV}$$

$$CO_2^+ + O \rightarrow O_2^+(v) + CO + 1.33 \text{eV}$$

Although the DR reaction branching ratios of vibrationally excited $O_2^+$ are different from those $O_2^+$ in the ground state, more than 99% of $O_2^+$ at lower altitudes, where most DR occur, are in the vibrational ground state. Thus for simplicity we ignore vibrationally excited $O_2^+$ and will consider them in future work.

Fig.3 shows the escape probability of O atoms with different kinetic energy as a function of altitude on modern Mars (XUV=1). Escape from altitude <190 km is negligible because of the high frequency of collisions. Above 190 km altitude, the escape probability increases slightly with kinetic energy between 2 and 4 eV and remains almost a constant > 4eV.

The calculated total escape rate of oxygen photochemical escape from modern Mars is $1.7\times10^{25}$ $s^{-1}$ under solar mean condition. The contribution of secondary energetic oxygen escape is only 0.2%. In comparison, escape rates of $1.6\times10^{25}$ $s^{-1}$ and $0.65\times10^{25}$ $s^{-1}$ are obtained in Fox and Hać (2014) for high and low solar activities respectively.

This result is consistent with the treatments of collisions in this work: constant $O-CO_2$ and $O-CO$ collision cross sections and small reduction of $O-O$ collision cross sections with energy (Tully and Johnson, 2001). Although our $O-CO_2$ and $O-CO$ collision cross sections are identical to those in Fox and Hać (2014), their energy loss depends on the scattering angle. In comparison, all $O-CO_2$ and $O-CO$ collisions lead to total energy loss in our work. Thus our model underestimates the escape probabilities of energetic oxygen to some degree. Because the main focus of this work is photochemical escape of oxygen from early Mars and the main background gas in the upper thermosphere is atomic species (Tian et al., 2008), our simple treatment is acceptable. In the future we will develop more realistic 3-D Monte Carlo model considering dependence of energy loss with scattering angle.

### 3 Early Mars Calculations

The model calculated primary oxygen photochemical escape rates are $1.7\times10^{25}$ $s^{-1}$, $8.4\times10^{25}$ $s^{-1}$, and $1.8\times10^{26}$ $s^{-1}$ for 1×, 3×, and 10× XUV respectively (Fig.4 b). These values are significantly lower compare with the results in Zhang et al. (1993) and Chassefière et al. (2013), but somewhat greater than those in Lammer et al. (2003). The calculated oxygen escape rate of $2.0\times10^{25}$ $s^{-1}$ in the 20× XUV case is in sharp contrast with the linear extrapolation of oxygen escape rates in the other works (Fig.4 b), despite the almost linear increase of oxygen DR rate with increasing solar XUV (Fig.4 a).

To understand these results, the altitude-dependent DR rates are plotted as functions of altitude-dependent escape probability in Fig.5. In the 1×, 3×, and 10× XUV cases, the DR rates decrease gradually with increasing escape probability because in most part of the upper atmospheres the $O_2^+$ number densities decrease with altitude (Fox and Hać, 2009) while escape probability increases with altitude. In the 20× XUV case, the decrease of DR rate with increasing escape probability is a much steeper function because of the existence of highly expanded atomic oxygen layers above the $O_2^+$ rich regions. At the altitude where the escape probability reaches 0.1 (200km, 260km, 400km, 800km for 1×, 3×, 10× and 20× XUV respectively), the DR rate in the 20× XUV case is more than 10 times lower than those in the 3× and 10× XUV cases, consistent with more rapid decrease of $CO_2$ density profile with altitude in this case (Tian et al., 2009).

The role of secondary energetic oxygen in photochemical oxygen escape from early Mars is more important for enhanced solar XUV flux. The column integrated formation efficiencies of secondary energetic oxygen atoms are 0.1%, 0.6%, 2% and 20% for 1×, 3×, 10× and 20× solar XUV situations respectively. Their contributions to escape rate are 0.2%, 1%, 5% and 30% of those of the primary energetic oxygen atoms, respectively. The high formation efficiency and contribution to total photochemical escape rate in the 20× XUV case is due to the decreased escape energy at high altitude, which leads to increased probability of secondary energetic oxygen atom production through collisions. The escape rates of secondary energetic oxygen atoms for the four solar XUV cases as functions of altitude are show in *Fig.5*b.

We further examined the contributions of third generation energetic oxygen atoms to total photochemical escape and found a <2% formation efficiency even in the 20× XUV case. Thus the contribution of $3^{rd}$ generation energetic oxygen atoms can be ignored.

Our results depend on the distributions of species in early Martian upper atmosphere calculated in Tian et al. (2009). The $O_2^+$ density profile in 1× XUV case in Tian et al. (2009) is overestimated in the lower thermosphere significantly when compared with that used in Fox and Hać (2009), which peaks at 130~140 km with a peak density of ~$10^5$ cm$^{-3}$ for both low and high solar activities. This is because $O_2^+$ was considered as a photochemical equilibrium species in Tian et al. (2009) and transport is important for dominant ion species such as $O_2^+$ in the ionospheres of Mars and Venus. In the middle to upper thermosphere, the $O_2^+$ density profile of Tian et al. (2009) is located between low and high solar activity when altitude higher than 170 km. This suggests that the vertical transport effect is important to determine the $O_2^+$ vertical distribution in the lower thermosphere but less important in the upper thermosphere. Because the escape probability of energetic oxygen atoms formed in the lower thermosphere (<180 km in the 1× XUV case) is nearly zero, the total escape rate calculated in this work is similar to that in previous works.

How the $O_2^+$ density profiles should be for early Mars and how vertical transport might have changed the $O_2^+$ density distributions and even the whole thermosphere structure from those calculated in Tian et al. (2009) are important questions requiring further study, but cannot be fully addressed in this work.

Observations of the upper atmospheres of close-in low mass exoplanets such as 55 Cnc e (Ehrenreich et al., 2012) could provide useful checks on the composition and thermal structures of these planets, which can greatly help our understanding the evolution of solar system terrestrial planets. We further note that our results are consistent with the opposite trends of $O_2^+$ DR rate and escape probability of energetic oxygen in different solar activity situation discussed in Fox and Hać (2009).

The time integrated oxygen loss through photochemical escape is ~$10^{43}$ oxygen atoms, equivalent to a global layer of 2 meters of water being lost from Mars since 4.5 Ga, assuming that all oxygen lost were from water. This value is significantly lower than those in previous studies: 30m water loss since 3.5 Ga in Zhang et al. (1993), 12m water loss since 3.5 Ga in Lammer et al. (2003), and 5m water loss since 4.1 Ga in Chassefière et al. (2013).

If our results are correct, because early Mars was exposed to more severe solar XUV radiation, photochemical escape should have been a less important loss mechanism of oxygen, and thereby water and $CO_2$, during the early Noachian. Because photochemical escape is one escape mechanism in which the escaping particles are not from the near-exobase region, our results would suggest that oxygen escape from early Noachian Mars could be closely associated with physics near the exobase. On the other hand, because photochemical escape is independent of the thermal structure of the upper atmosphere, the insignificance of photochemical escape in planetary atmospheres under strong XUV radiation would lend support to the recently proposed total conservation of atmosphere escape hypothesis (Tian et al., 2013), in which atmosphere escape from intensively XUV irradiated and thus highly expanded planetary upper atmospheres is controlled by the thermal structure and energy consumption in planetary upper atmosphere instead of the physical processes around the exobase level.

Acknowledgement: The authors are supported by the National Natural Science Foundation of China (41175039) and the Startup Fund of the Ministry of Education of China. The authors thank two anonymous reviewers who provided constructive suggestions which improved the quality of this paper.

## *References*

Chassefière, E., Langlais, B., Quesnel, Y., Leblanc, F., 2013. The fate of early Mars' lost water: The role of serpentinization. Journal of Geophysical Research: Planets. 118, 1123-1134.
http://dx.doi.org/10.1002/jgre.20089.

Cipriani, F., Leblanc, F., Berthelier, J. J., 2007. Martian corona: Nonthermal sources of hot heavy species. Journal of Geophysical Research-Planets. 112.
http://dx.doi.org/10.1029/2006je002818.

Ehrenreich, D., et al., 2012. Hint of a transiting extended atmosphere on 55 Cancri b. Astronomy & Astrophysics. 547.
http://dx.doi.org/10.1051/0004-6361/201219981.

Fox, J. L., 1993. On the Escape of Oxygen and Hydrogen from Mars. Geophysical Research Letters. 20, 1747-1750.
http://dx.doi.org/10.1029/93gl01118.

Fox, J. L., Hać, A. B., 2009. Photochemical escape of oxygen from Mars: A comparison of the exobase approximation to a Monte Carlo method. Icarus. 204, 527-544.
http://dx.doi.org/10.1016/j.icarus.2009.07.005.

Fox, J. L., Hać, A. B., 2014. The escape of O from Mars: Sensitivity to the elastic cross sections. Icarus. 228, 375-385.
http://dx.doi.org/10.1016/j.icarus.2013.10.014.

Gröller, H., Lichtenegger, H., Lammer, H., Shematovich, V. I., 2014. Hot oxygen and carbon escape from the martian atmosphere. Planetary and Space Science. 98, 93-105.
http://dx.doi.org/10.1016/j.pss.2014.01.007.

Hodges, R. R., 2000. Distributions of hot oxygen for Venus and Mars. Journal of Geophysical Research-Planets. 105, 6971-6981.
http://dx.doi.org/10.1029/1999je001138.

Johnson, R. E., et al., 2008. Exospheres and Atmospheric Escape. Space Science Reviews. 139, 355-397.
http://dx.doi.org/10.1007/s11214-008-9415-3.

Kharchenko, V., Dalgarno, A., Zygelman, B., Yee, J. H., 2000. Energy transfer in collisions of oxygen atoms in the terrestrial atmosphere. Journal of Geophysical Research-Space Physics. 105, 24899-24906.
http://dx.doi.org/10.1029/2000ja000085.

Kim, J., Nagy, A. F., Fox, J. L., Cravens, T. E., 1998. Solar cycle variability of hot oxygen atoms at Mars. Journal of Geophysical Research-Space Physics. 103, 29339-29342.
http://dx.doi.org/10.1029/98ja02727.

Krestyanikova, M. A., Shematovich, V. I., 2006. Stochastic models of hot planetary and satellite coronas: A hot oxygen corona of Mars. Solar System Research. 40, 384-392.
http://dx.doi.org/10.1134/S0038094606050030.

Lammer, H., Bauer, S. J., 1991. Nonthermal Atmospheric Escape from Mars and Titan. Journal of Geophysical Research-Space Physics. 96, 1819-1825.
http://dx.doi.org/10.1029/90ja01676.

Lammer, H., Kasting, J. F., Chassefiere, E., Johnson, R. E., Kulikov, Y. N., Tian, F., 2008. Atmospheric Escape and Evolution of Terrestrial Planets and Satellites. Space Science Reviews. 139, 399-436.
http://dx.doi.org/10.1007/s11214-008-9413-5.

Lammer, H., et al., 2003. Loss of water from Mars: Implications for the oxidation of the soil. Icarus. 165, 9-25.
http://dx.doi.org/10.1016/S0019-1035(03)00170-2.

Shematovich, V. I., 2013. Suprathermal oxygen and hydrogen atoms in the upper Martian atmosphere. Solar System Research. 47, 437-445.
http://dx.doi.org/10.1134/S0038094613060087.

Shizgal, B. D., Arkos, G. G., 1996. Nonthermal escape of the atmospheres of Venus, Earth, and Mars. Reviews of Geophysics. 34, 483-505.
http://dx.doi.org/10.1029/96rg02213.

Tian, F., Chassefière, E., Leblanc, F., Brain, D., 2013 Atmospheric Escape and Climate Evolution of Terrestrial Planets. In: S. J. Mackwell, (Ed.), Comparative Climatology of Terrestrial Planets, pp. 567-581.
http://dx.doi.org/10.2458/azu_uapress_9780816530595-ch23.

Tian, F., Kasting, J. F., Liu, H. L., Roble, R. G., 2008. Hydrodynamic planetary thermosphere model: 1. Response of the Earth's thermosphere to extreme solar EUV conditions and the significance of adiabatic cooling. Journal of Geophysical Research-Planets. 113, E05008.
http://dx.doi.org/10.1029/2007je002946.

Tian, F., Kasting, J. F., Solomon, S. C., 2009. Thermal escape of carbon from the early Martian atmosphere. Geophysical Research Letters. 36, L02205.

http://dx.doi.org/10.1029/2008gl036513.

Tully, C., Johnson, R. E., 2001. Low energy collisions between ground-state oxygen atoms. Planetary and Space Science. 49, 533-537.
http://dx.doi.org/10.1016/S0032-0633(01)00002-2.

Valeille, A., Combi, M. R., Tenishev, V., Bougher, S. W., Nagy, A. F., 2010. A study of suprathermal oxygen atoms in Mars upper thermosphere and exosphere over the range of limiting conditions. Icarus. 206, 18-27.
http://dx.doi.org/10.1016/j.icarus.2008.08.018.

Yagi, M., Leblanc, F., Chaufray, J. Y., Gonzalez-Galindo, F., Hess, S., Modolo, R., 2012. Mars exospheric thermal and non-thermal components: Seasonal and local variations. Icarus. 221, 682-693.
http://dx.doi.org/10.1016/j.icarus.2012.07.022.

Zhang, M. H. G., Luhmann, J. G., Bougher, S. W., Nagy, A. F., 1993. The Ancient Oxygen Exosphere of Mars - Implications for Atmosphere Evolution. Journal of Geophysical Research-Planets. 98, 10915-10923.
http://dx.doi.org/10.1029/93je00231.

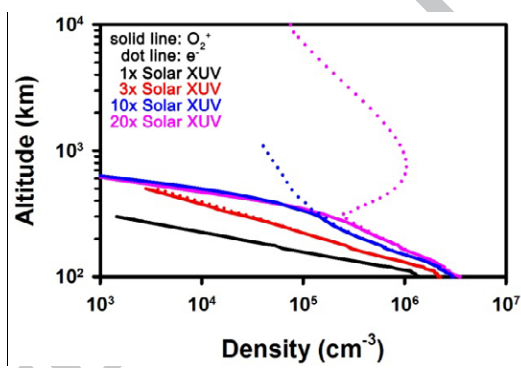

*Fig.1. Density distributions of $O_2^+$ (solid curves) and $e^-$ (dot curves) in 1× (black), 3× (red), 10× (blue), 20× (purple) solar XUV cases.*

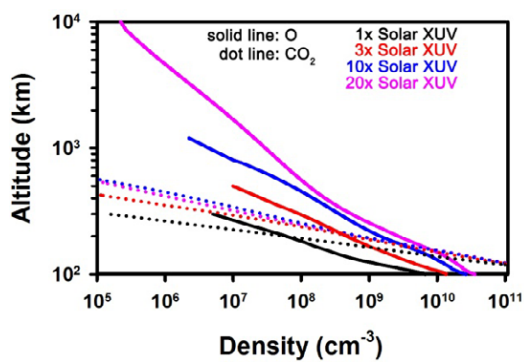

*Fig.2. Density profiles of O (solid curves) and $CO_2$ (dotted curves) in different solar XUV level.*

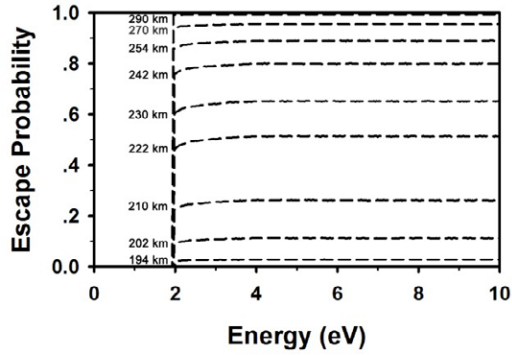

*Fig.3. Energy dependent O escape probabilities from modern Mars.*

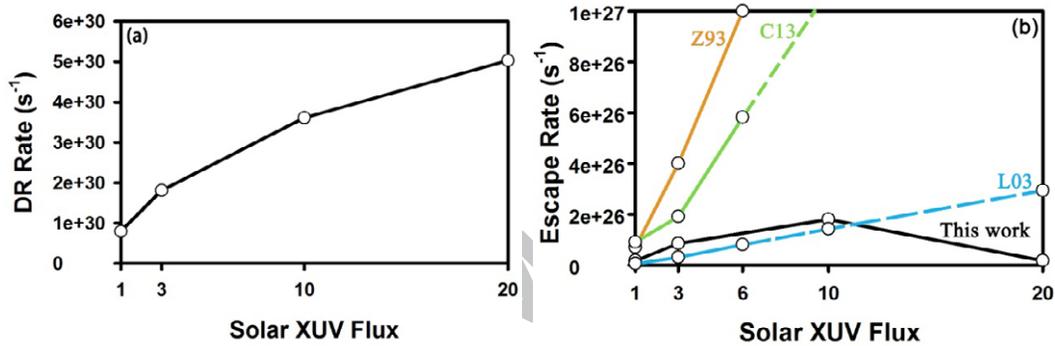

*Fig.4. Model calculated oxygen DR rates (panel a) and oxygen escape rate (panel b) under different solar XUV conditions. Z93=Zhang et al. (1993), C13=Chassefière et al. (2013), L03=Lammer et al. (2003) and black solid line represents this work. These works calculated escape rates at 1, 3 and 6 times present solar XUV flux, and linear extrapolations (marked by dashed lines) are used to obtain the escape rates at >6× present solar XUV cases.*

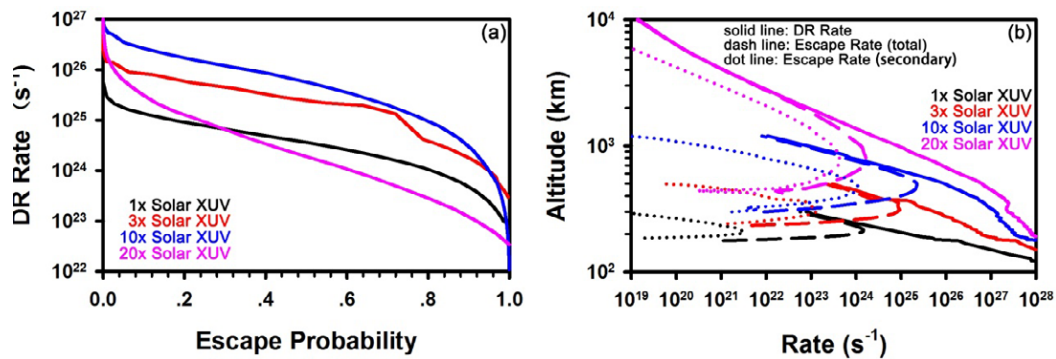

*Fig.5.(a) DR Rates as functions of escape probability in different solar XUV cases; (b) DR rates, total escape rates and secondary escape rates (solid, dashed, and dotted curves respectively) as functions of altitude.*

Highlights

- A 1-D Monte-Carlo Model is developed to calculate oxygen photochemical escape from early Mars.

- On early Mars the escape rate of energetic oxygen atoms produced from $O_2^+$ dissociative recombination does not increase monotonically with increasing solar XUV flux. The maximum photochemical escape rate was reached between 3.5 and 4.2 Ga.

- Photochemical escape should be less important than previously thought for the loss of water and/or $CO_2$ from early Mars. The photochemical escape of oxygen alone accounts for less than 2 meters global equivalent layer of water over the entire evolution history of Mars.